\documentclass{elsart}

 \usepackage{graphics}
 \usepackage{graphicx}

\usepackage{amssymb}

\usepackage{multirow}

\begin{document}

\def\KHeXLaEneVal{$ 6466.7 \pm 2.5$}
\def\KHeXLbEneVal{$ 8723.3 \pm 4.6$}
\def\KHeXLgEneVal{$ 9760.1 \pm 7.7$}
\def\KHeXLdEneVal{$10328.7 \pm 35.1$}
\def\KHeXLaEneValRoundOut{$ 6467 \pm  3$}
\def\KHeXShift{$2 \pm 2$}
\def\KHeXSystErr{$\pm 2$}

\begin{frontmatter}



\title{Precision measurement of the $3d \to 2p$ x-ray energy
 in kaonic $^4$He}

 
 \author[riken]{S. Okada},
 \ead{sokada@riken.jp}
 \author[victoria]{G. Beer},
 \author[seoul]{H. Bhang},
 \author[smi]{M. Cargnelli},
 \author[tus]{J. Chiba},
 \author[seoul]{Seonho Choi},
 \author[infn]{C. Curceanu},
 \author[titech]{Y. Fukuda},
 \author[tus]{T. Hanaki},
 \author[ut]{R. S. Hayano},
 \author[riken]{M. Iio},
 \author[ut]{T. Ishikawa},
 \author[kek]{S. Ishimoto},
 \author[smi]{T. Ishiwatari},
 \author[riken]{K. Itahashi},
 \author[kek]{M. Iwai},
 \author[riken,titech]{M. Iwasaki},
 \author[smi]{B. Juh$\acute{\mbox{a}}$sz},
 \author[smi,tum]{P. Kienle},
 \author[smi]{J. Marton},
 \author[riken]{Y. Matsuda},
 \author[riken]{H. Ohnishi},
 \author[riken]{H. Outa},
 \author[titech]{M. Sato\thanksref{now}},
 \author[smi]{P. Schmid},
 \author[kek]{S. Suzuki},
 \author[riken]{T. Suzuki},
 \author[ut]{H. Tatsuno},
 \author[riken]{D. Tomono},
 \author[smi]{E. Widmann},
 \author[riken,ut]{T. Yamazaki},
 \author[seoul]{H. Yim},
 \author[smi]{J. Zmeskal}

 \address[riken]{RIKEN Nishina Center, RIKEN, Saitama 351-0198, Japan}
 \address[victoria]{Department of Physics and Astronomy, University of
 Victoria, British Columbia V8W 3P6, Canada}
 \address[seoul]{Department of Physics, Seoul National University,
 Seoul 151-742, South Korea}
 \address[smi]{Stefan Meyer Institut f$\ddot{\mbox{u}}$r subatomare
 Physik, Austrian Academy of Sciences, A-1090 Vienna, Austria}
 \address[tus]{Department of Physics, Tokyo University of Science,
 Chiba 278-8510, Japan}
 \address[infn]{Laboratori Nazionali di Frascati, INFN,
 I-00044 Frascati, Italy}
 \address[titech]{Department of Physics, Tokyo Institute of Technology,
 Tokyo 152-8551, Japan}
 \address[ut]{Department of Physics, The University of Tokyo,
 Tokyo 113-0033, Japan}
 \address[kek]{High Energy Accelerator Research Organization (KEK),
 Ibaraki 305-0801, Japan}
 \address[tum]{Physik Department, Technische Universit$\ddot{\mbox{a}}$t
 M$\ddot{\mbox{u}}$nchen, D-85748 Garching, Germany}
 \thanks[now]{Present address: RIKEN Nishina Center, RIKEN,
 Saitama 351-0198, Japan}

\begin{abstract}
 We have measured the Balmer-series x-rays of kaonic $^4$He atoms using
 novel large-area silicon drift x-ray detectors in order to study the
 low-energy  $\bar{K}$-nucleus strong interaction. The energy of the
 $3d \to 2p$ transition was determined to be
 \KHeXLaEneValRoundOut\ (stat)  \KHeXSystErr\ (syst) eV.
 The resulting strong-interaction
 energy-level shift is in agreement with
 theoretical calculations, thus eliminating a long-standing discrepancy
 between theory and experiment.
\end{abstract}

\begin{keyword}
 kaonic atom \sep
 x-ray spectroscopy \sep
 silicon drift detector
\PACS
 13.75.Jz \sep
 25.80.Nv \sep
 36.10.Gv
\end{keyword}
\end{frontmatter}

\section{Introduction}
\label{section:introduction}
The measurement of the strong-interaction energy-level shift and width
of kaonic atom x-rays offers a unique possibility to precisely
determine the $\bar{K}$-nucleus strong interaction in the low energy
limit.
Therefore many experiments have been performed to collect data on
a variety of targets from hydrogen to uranium.
It has been known that most of the available kaonic-atom data can be
fitted fairly well for $Z \ge 2$ by optical-potential models \cite{Bat97}
with the exception of kaonic helium and oxygen.

The strong-interaction shift of the
$2p$ level $\Delta E_{2p}$ for kaonic $^4$He
has been previously measured in three experiments.
Note that
$\Delta E_{2p}$ is defined
as $\Delta E_{2p} \equiv - (E_{(2,p)} - E_{(2,p)}^{EM})$,
where $E_{(n,l)}$ is the energy 
of the level with principal quantum number $n$ and the orbital angular
momentum $l$,
and $E_{(n,l)}^{EM}$ is the energy calculated using only
the electromagnetic interaction (EM).
The average of the three previous results gives
$\Delta E_{2p} = -43 \pm 8$ eV \cite{Wie71,Bat79,Bai83},
while most of the theoretical calculations
give $\Delta E_{2p} \sim 0$ eV \cite{Bat90,Hir00,FriPC}
({\it e.g.} $\Delta E_{2p} = -0.13 \pm 0.02$ eV \cite{Bat90}).
They disagree by more than five standard deviations, and
this discrepancy is known as the ``kaonic helium puzzle''.
Therefore an accurate re-measurement of the energy shift
of the $2p$ level of kaonic helium has been long awaited.

In the present experiment
we have performed a measurement of the Balmer-series x-rays of
kaonic $^4$He atoms,
setting as our experimental objective a precision of
$\sim$ 2 eV, thus shedding new light on the kaonic helium puzzle.

\section{Experiment}
\label{section:experiment}

The experiment E570 was carried out at the K5 beamline of the
KEK 12-GeV proton synchrotron (PS).
We accumulated data in two periods -- 520 hours in October 2005 (cycle 1)
and 260 hours in December 2005 (cycle 2).
The experimental apparatus was essentially the same as that of the former
KEK-PS E549 experiment \cite{Sat07} performed at the same beamline,
except for the inclusion of x-ray detectors and energy calibration
foils in the helium-target cryostat.
A schematic view of the E570 setup around the target is shown in
Fig. \ref{fig:setup}.
The detailed description of the experimental setup
is given in a separate paper \cite{Sat07}.

The kaonic $^4$He atoms were produced by means of the stopped-$K^-$
reaction using a superfluid $^4$He target
(cylindrical shape
15 cm long and  20 cm in diameter at a density of 0.145 g/cm$^3$).
Incident negatively-charged kaons with momentum $\sim$650 MeV/$c$
were degraded in carbon degraders,
counted with beamline counters,
tracked by a high-rate beamline drift chamber
and stopped inside the $^4$He target.
The energy losses before stopping
were measured in a set of scintillation counters, T0.
X-rays emitted from the kaonic $^4$He atoms
were detected by eight x-ray detectors which
viewed the target from downstream through the 75 $\mu$m-thick Mylar
window of the target vessel.
Secondary charged particles produced in the kaon-absorption process
following emission of kaonic x-rays were detected by charged-particle
trigger/tracking systems
placed on the left, right, top and bottom of the target.

In the present experiment, a significant improvement over the past
experiments
was achieved by incorporating the following features:

\begin{figure}
 \begin{center}
  \includegraphics*[width=12cm]{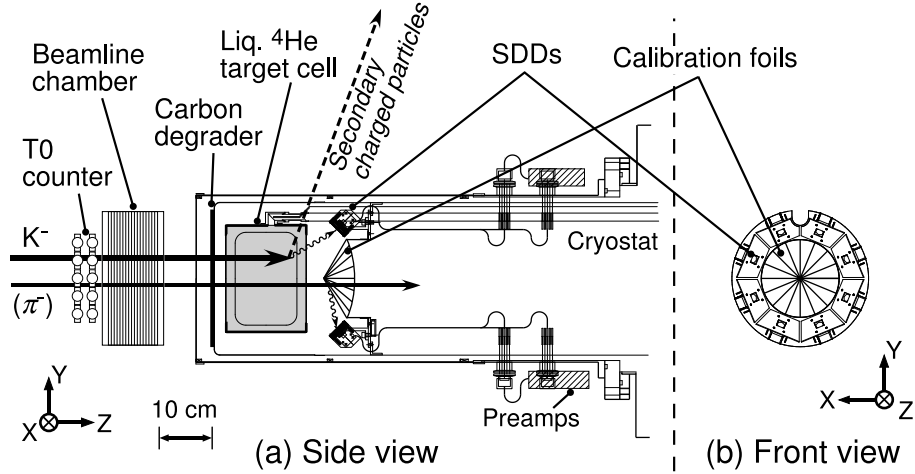}
 \end{center}
  \caption{(a) A schematic side view of
 the E570 setup around the cylindrical target
 with the x-ray detection system.
 (b) A front view of the silicon drift detector (SDD) assembly.
 Eight SDDs are mounted on holders tilted at a 45 degree angle
 to the beam center in an annular-shaped pattern.
 Fan-shaped high-purity titanium and nickel foils are put
 alternately on a cone-shaped support located on the beam axis.}
 \label{fig:setup}
\end{figure}

\subsection{Silicon drift detectors}

As x-ray detectors, we employed eight silicon drift detectors (SDDs)
produced by KETEK GmbH \cite{KETEK}, each having an effective area
of 100 mm$^2$ and a 260 $\mu$m-thick active layer
with an energy resolution of $\sim$190
eV (FWHM) at 6.4 keV,
which corresponds to the kaonic-helium $3d \to 2p$ x-ray energy.
The temperature of the SDDs was kept at $\sim$83 K during the experiment
by a connection to the thermal-radiation shield for the helium target
cooled by liquid nitrogen.

In the SDD, the electrons produced by an x-ray hit drift radially
toward the central anode where they are collected.
The small anode size (and hence small capacitance) is essential to
realize the good energy resolution despite the large effective area.
The energy resolution is about twice as good as that of the Si(Li)
x-ray detectors used in the previous three experiments.
The time resolution is comparable with that of a Si(Li) detector.

The small anode area also makes it possible to reduce the active layer
thickness, while the capacitance is still kept small.
The thin active layer of 260 $\mu$m
(less than 1/10th of the previously used detectors)
helps to reduce continuum background caused by the
Compton scattering occurring inside the detector.

\subsection{Cuts applied to reduce background}

We required that the reaction vertices reconstructed from
an incident kaon track and an outgoing secondary charged particle track
should be within the target,
which is called the ``fiducial volume cut''.
Moreover, in-flight kaon decay/reaction events were rejected
by applying a correlation cut between the $z$-coordinate of the reaction
vertex and the energy loss in T0.
As a result,
continuum background events were drastically reduced.

\subsection{In-beam energy calibration}

The energy calibration was done by using characteristic x-rays induced by 
charged particles ($i.e.~ \pi^-$, which abundantly existed
in the incident beam)
on high-purity titanium and nickel foils placed
just behind the target cell.
The energy of the kaonic-helium $3d \to 2p$ x-ray,
$\sim$6.4 keV, lies between the characteristic x-ray energies,
4.5 keV(Ti) and 7.5 keV(Ni).
To obtain high-statistics energy calibration spectra,
we accumulated SDD self-triggered events together
with the stopped-$K^-$ triggered events,
which provide high-accuracy in-situ calibration spectra.

To avoid detecting the background characteristic x-rays
from other than the titanium and nickel,
high-purity aluminum foils were placed on all objects
in the view of the SDDs.

\section{Analysis}
\label{section:analysis}

\begin{figure}
 \begin{center}
  \includegraphics*[width=10cm]{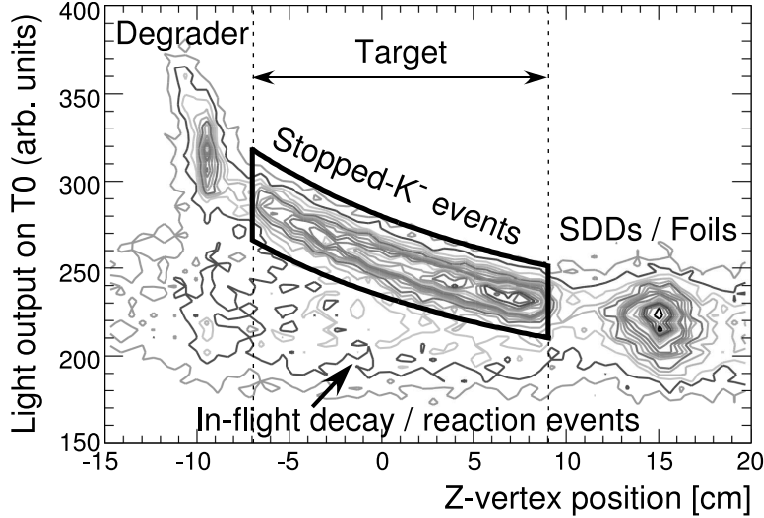}
 \end{center}
 \caption{A typical density plot between
 the $z$-coordinate of the reaction vertex and
 the light output on T0, used to reject in-flight kaon
 decay/reaction events.}
 \label{fig:t0verz}
\end{figure}

Figure\ \ref{fig:t0verz} shows the correlation between the $z$-coordinate
of the reaction vertex and the light output of T0.
Each component of the target assembly
(a carbon degrader, a target cell and SDDs/foils) is clearly seen.
We applied a fiducial volume cut
of $-7.0 < z < 9.0$ cm on the $z$ coordinate as shown
in Fig.\ \ref{fig:t0verz},
and of $\sqrt{x^2+y^2} < 11.0$ cm on the radius from the target center.
Slower incident kaons, which give larger light output on T0,
stop upstream in the target,
while faster kaons (hence smaller pulse height) stop downstream.
Events which follow this trend were selected
as stopped-$K^-$ events when lying within the solid-lined box
in Fig. \ref{fig:t0verz}.

Stopped-$K^-$-timing events were selected using SDD timing information
to reduce the accidental background.
Time resolution of the SDD after time-walk correction
was $\sim$ 160 ns ($\sigma$) at $\sim$83 K,
which reflected the drift-time distribution of the electrons
in the SDD.
Data within $\pm$ 2 standard deviations from the average SDD hit
timing were selected.

\begin{figure}
 \begin{center}
  \includegraphics*[width=10cm]{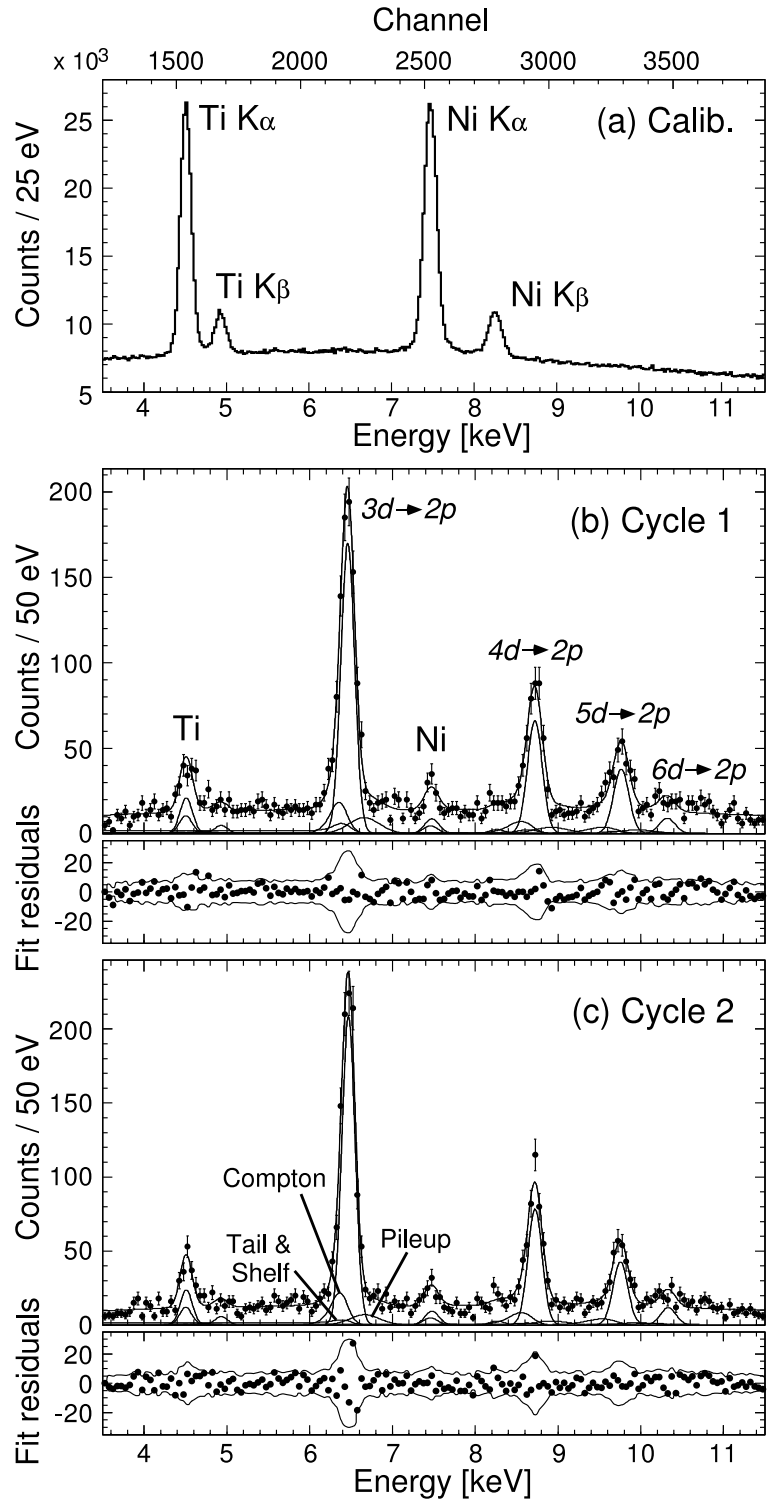}
 \end{center}
 \caption{(a) A typical x-ray spectrum for self-triggered events
 which provides high-statistics energy-calibration information.
 (b)(c) Measured x-ray spectra for stopped-$K^-$ events obtained
 from the runs in October 2005 (cycle 1) and December 2005
 (cycle 2) respectively.
 A fit line is also shown for each spectrum,
 along with individual functions of the fit.
 The fit residuals are shown under each spectrum,
 with thin lines denoting the $\pm 2\sigma$ values of the data,
 where $\sigma$ is the standard deviation due to the counting
 statistics.}
 \label{fig:spectra}
\end{figure}

Figure\ \ref{fig:spectra} (a) shows a typical x-ray spectrum
for SDD self-triggered events,
which is used for the energy calibration.
Characteristic x-ray peaks of titanium and nickel were obtained
with high statistics.
Typical yields of titanium K$_\alpha$ peaks
are 5 $\times 10^2$ events per hour for each SDD.
Time-dependent gain drift was corrected about every 20 hours.
The energy scale was calibrated
by K$_\alpha$ lines of titanium and nickel
with the well-known energies \cite{NIST}
and intensity ratios \cite{XDB}
of K$_{\alpha 1}$ and K$_{\alpha 2}$.

After applying the event selections described above
and calibrating the energy scale,
we obtained x-ray energy spectra for stopped-$K^-$ triggered events
shown in Fig.\ \ref{fig:spectra}.
Kaonic-helium $3d \to 2p$, $4d \to 2p$ and $5d \to 2p$
transitions are clearly observed,
while the Ti and Ni x-ray peaks are greatly suppressed.
Figure\ \ref{fig:spectra} (b) and (c) respectively show
the x-ray spectra taken in the runs
in October 2005 (cycle 1) and December 2005 (cycle 2).
In cycle 1, only 3 out of 8 SDDs yielded useful data;
the faulty detectors were then replaced, and 7 SDDs were functional
in cycle 2.
In total,
$\sim 7 \times 10^2$ (cycle 1) and $\sim 8 \times 10^2$ (cycle 2) events
of $3d \to 2p$ x-rays
were accumulated for each cycle.
The total yields are approximately equivalent for both runs
since the beam time of the October run was twice that of 
the December run.
In comparison to the most recent measurement
of the kaonic $^4$He spectrum \cite{Bai83}, 
we achieved $\sim$2 times better energy resolution,
$\sim$3 times higher statistics,
and $\sim$6 times better signal-to-noise ratio.

\section{Spectral fitting and results}
\label{section:fitting}

During the course of the present analysis of spectra from SDDs, many
fine details of the signal and calibration pulses were found necessary
to attain eV accuracy and will be discussed extensively in a paper to
follow.
Here, we briefly summarize the spectral fitting method
with a SDD-response function studied in detail,
and show the fit results.

For fitting the spectra of kaonic-helium x-ray peaks,
a convolution of a Gaussian (representing the detector response)
and a Lorentz function (natural width),
``Voigt function'', was adopted as the main-peak function,
whereas a Gaussian response was employed as a main-peak function
for fitting the characteristic x-ray peaks
since their natural width is much less than the energy
resolution of the SDD.

An energy-dependent experimental energy resolution was employed
as is usually the case for silicon detectors:
$\Delta E \mbox{(FWHM)} = 2.35 \omega \sqrt{ W_N^2 + FE/\omega}$,
where $W_N$ denotes the contribution of noise to the resolution
(independent of the x-ray energy), $E$ is the x-ray energy,
$F$ is the Fano factor ($\approx$ 0.12 for silicon), and
$\omega$ is the average energy for electron-hole creation in silicon.
Here, $\omega$ was fixed to be 3.81 eV,
and $F$ and $W_N$ were introduced as free fit-parameters
for the self-triggered-event spectra.

Because of the large incoherent (Compton) total scattering cross section
of liquid $^4$He ($\sim 1$ barn/atom at photon energies of
$\sim$10 keV), a low-energy tail structure
due to the Compton scattering must be taken into account.
The convolution of an exponential function with a Gaussian
was adopted as the spectral shape (called the ``Compton tail function'').
All parameters of the Compton tail function relative to the main peak
were estimated by fitting the simulated x-ray energy spectra smeared with a
Gaussian resolution function.
The x-ray spectra were simulated with the GEANT code
using the Low Energy Compton Scattering (LECS) package \cite{LECS}
with a realistic setup of E570 and the measured stopped-$K^-$
distribution.

We also have waveform data available from flash ADCs (FADCs),
which were accumulated as well as ordinary
comparator-type pulse-height ADC data,
although the FADC data are available only for about half of cycle 1.
Using the waveform analysis,
it is shown that there is a non-negligible pileup effect
due to the high-rate beam condition of E570.
The spectral function attributed to the pileup events could be estimated
as a Gaussian on the right flank of the main-peak function
(called the ``pileup Gaussian'').
The mean and sigma parameters of the pileup Gaussian were estimated
using FADC data.

There are many empirical investigations of the response function of
silicon detectors using monoenergetic x-rays
($e.g.$ \cite{ResponseFunc01,ResponseFunc03}),
and it is known that there is an exponential-like feature decreasing
steeply in intensity towards lower energy on the left flank
of the main-peak function (called the ``tail function'')
and a flat shelf-like feature which extends to near zero energy
(called the ``shelf function'') \cite{ResponseFunc03},
due to electron transport processes and
imperfections in the fabrication processes.
These effects were also taken into account
in the spectral fitting separately from the Compton tailing effect
in the liquid $^4$He target mentioned above.

The intensity ratios of these components
(pileup Gaussian, tail and shelf functions) to
the main-peak component 
were estimated by fitting
the high-statistics x-ray spectra for self-triggered events.
The estimated intensity ratios were then fixed
in the fits of the x-ray spectra for stopped-$K^-$ triggered events.

Resulting fit-lines are overlaid
on the x-ray spectra shown in Fig.\ \ref{fig:spectra} (b) and (c)
with each contribution:
main Voigtian, Compton tail function,
pileup Gaussian, tail function and shelf function.
The fit residuals are also shown under each spectrum,
with thin lines denoting the $\pm 2\sigma$ values of the data,
where $\sigma$ is the standard deviation due to the counting
statistics.
In the fits, the intensity and mean parameters
for each kaonic helium x-ray transition are independent.
As a result,
the kaonic $^4$He x-ray energy of the $3d \to 2p$ transition
was determined to be
\begin{eqnarray}
 E_{(3,d)} - E_{(2,p)} = \mbox{\KHeXLaEneValRoundOut\ (stat)
  \KHeXSystErr\ (syst) ~~eV}
  \label{eqn:3d2pXrayEnergy}
\end{eqnarray}
where the first error is statistical and the second is systematic.
The quoted systematic error is a linear summation of the contributions
from the intensity ambiguities of
the Compton tail , pileup, shelf and tail functions
for kaonic-helium x-rays.
The other transition energies ($4d \to 2p$ and $5d \to 2p$)
obtained in this fit are
listed in table \ref{table:transition-energy} 
with only statistical errors.
In this table, we also tabulate
the EM values updated from Refs. \cite{Wie71,Bat79,Bai83}
by Koike \cite{KoiPC}
using the latest kaon mass given by
the particle data group (PDG) \cite{PDG}.
These values are consistent with another recent calculation
\cite{San05} and differ slightly from the ones used
in previous experiments \cite{Wie71,Bat79,Bai83}.
The intrinsic width obtained in the fits
seems to be very small and it needs more study
to disentangle it from instrumental effects.

\begin{table}[hbt]
 \begin{center}
  \begin{tabular}{c|c|c|c} \hline \hline
   Transition & $3d \to 2p$ & $4d \to 2p$ & $5d \to 2p$ \\
   \hline
   \hline
   Measured energy (eV) &
   \KHeXLaEneVal\  &
   \KHeXLbEneVal\  &
   \KHeXLgEneVal\  \\
   \hline
   EM calc. energy (eV) \cite{KoiPC} &
   6463.5 & 8721.7 & 9766.8 \\
   \hline \hline
  \end{tabular}
  \caption{Measured and EM calculated \cite{KoiPC}
  kaonic $^4$He x-ray energies of 
  $3d \to 2p$, $4d \to 2p$ and $5d \to 2p$ transitions.
  The quoted error is purely statistical.}
  \label{table:transition-energy}
 \end{center}
\end{table}

Since the strong-interaction shifts are negligibly small
for the levels with the principal quantum number $n$ larger than two,
the $2p$-level shift $\Delta E_{2p}$ can be derived from the
Balmer-series x-ray energies using the equation:
\begin{eqnarray}
 \Delta E_{2p} = (E_{(n,d)}-E_{(2,p)}) - (E_{(n,d)}^{EM}-E_{(2,p)}^{EM}),
  \label{eq:deltaE}
\end{eqnarray}
where $E_{(n,d)}-E_{(2,p)}$ and $E_{(n,d)}^{EM}-E_{(2,p)}^{EM}$
correspond to the measured and calculated EM x-ray energies, respectively.
To combine all statistics of the observed Balmer-series lines,
we calculated $\Delta E_{2p}$ for each line
using Eq. \ref{eq:deltaE} and took their statistical averages;
the $2p$-level shift was then derived as
\begin{eqnarray}
 \Delta E_{2p} = \mbox{\KHeXShift\ (stat)
  \KHeXSystErr\ (syst) eV,}
\end{eqnarray}
where EM values listed in table \ref{table:transition-energy}
\cite{KoiPC} were adopted,
and the systematic error was estimated in the way mentioned above.
Note that the EM energy and thus
the energy shift are sensitive to the value of the kaon mass,
for which two slightly disagreeing measurements exist \cite{PDG}
leading to a large error in the PDG value.
If the kaon mass changes by one standard
deviation from the current value of 493.677(16) MeV/$c^2$ \cite{PDG},
$\Delta E_{2p}$ changes by about 0.2 eV \cite{KoiPC}.

\section{Conclusion}
\label{section:conclusion}

In conclusion,
we have measured the Balmer-series x-rays of kaonic $^4$He atoms
using silicon drift detectors
which lead to a much improved energy resolution and
signal-to-noise ratio compared to the Si(Li) x-ray
detectors used in the past experiments.
The kaonic $^4$He x-ray energy of the $3d \to 2p$ transition
was determined to be \KHeXLaEneValRoundOut\ (stat)  \KHeXSystErr\ (syst) eV.

\begin{figure}
 \begin{center}
  \includegraphics*[width=10cm]{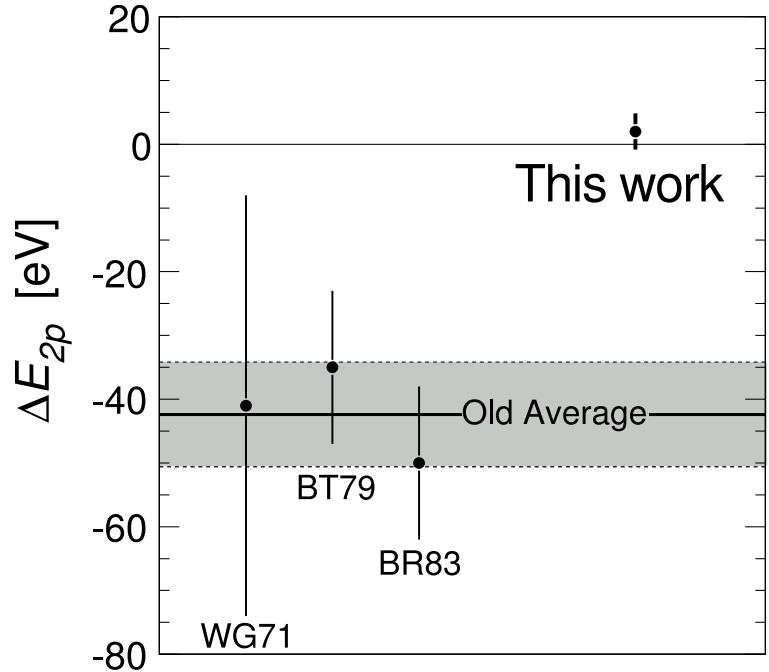}
 \end{center}
 \caption{The $2p$-level shift of kaonic $^4$He,
 $\Delta E_{2p}$, obtained from this
 work and the past three experiments 
 (WG71 \cite{Wie71}, BT79 \cite{Bat79}, BR83 \cite{Bai83}).
 Error bars show quadratically added statistical and systematic
 errors.
 The average of these past experiments
 is indicated by the horizontal gray band.}
 \label{fig:comparison}
\end{figure}

Using three observed transition lines
($3d \to 2p$, $4d \to 2p$ and $5d \to 2p$)
with the corresponding EM values \cite{KoiPC},
the $2p$-level shift was deduced as
$\Delta E_{2p} = $ \KHeXShift\ (stat) \KHeXSystErr\ (syst) eV.
Figure \ref{fig:comparison} shows a comparison
of the $2p$-level shifts between this work
and the previous experiments \cite{Wie71,Bat79,Bai83}.
Our result excludes the earlier claim of a large shift of about $-40$ eV.

The theoretical calculations of the shift are very close to zero
($\sim -0.1$) eV
by an analysis with global fits to existing kaonic-atom
x-ray data on various nuclei using an optical potential \cite{Bat97,Bat90},
and also by a calculation using an SU(3) chiral unitary model \cite{Hir00}.
A recent calculation by Friedman gives a value
of $-0.4$ eV as the lowest possible one \cite{FriPC},
when the non-linear density dependence is included \cite{Mar06}.
On the other hand,
Akaishi calculated the shift as a function of the real part
($U_0$) of the $\bar{K} N$ potential depth at
a certain coupled potential depth ($U_{coupl} = 120$ MeV) \cite{Aka05}.
The calculation was based on the coupled-channel approach between
the $\bar{K} N$ channel and the $\Sigma ~\pi$ decay channel.
A large shift ($|\Delta E_{2p}| \sim 10$ eV) is predicted near the
resonance between atomic and nuclear poles, when the potential depth
is at around $\sim$ 200 MeV.
The presently observed small shift disfavors
the values of ($U_0, U_{coupl}$) = ($\sim$ 200 MeV, 120 MeV)
within his framework.

Our careful and precise determination of the $2p$-level shift 
resolved the long-standing kaonic helium puzzle.
The present data alone are not sufficient to deduce
the $\bar{K}$-nucleus potential strength at the center of the nucleus.
A unified study with the $2p$ width
to be determined in further analysis 
and with data to be
collected in kaonic $^3$He x-ray spectroscopy \cite{E17} will
indubitably yield invaluable constraints for the theories.

 \section*{Acknowledgments}
 \label{section:acknowledgments}

 We wish to thank Y.~Akaishi for theoretical discussions.
 We acknowledge T.~Koike who made an electromagnetic calculation
 of the kaonic helium atomic states.
 We are grateful to the KEK staff members of 
 the beam channel group, accelerator group, cryogenics group,
 and electronics group, for support of the present experiment.
 We are indebted to H.~Schneider for his technical support
 in connection with the SDDs.
 We also owe much to T.~Taniguchi for his contribution in
 developing electronics.
 We would like to thank the PSI staff
 for their help during a test experiment.
 This research was
 supported by KEK, RIKEN, SMI and
 Grant-in-Aid for Scientific Research
 from the Ministry of Education of Japan,
 No. 17340087, No. 14102005 and No. 17070007.




\begin{thebibliography}{00}





 \bibitem{Bat97} C.~J.~Batty, E.~Friedman and A.~Gal,
	 Phys. Reports {\bf 287}, 385 (1997).

 \bibitem{Wie71} C.~E.~Wiegand and R.~Pehl,
	 Phys. Rev. Lett. {\bf 27}, 1410 (1971).
	 
 \bibitem{Bat79} C.~J.~Batty {\em et al.},
	 Nucl. Phys. {\bf A 326}, 455 (1979).
	 
 \bibitem{Bai83} S.~Baird {\em et al.},
	 Nucl. Phys. {\bf A 392}, 297 (1983).
	 
 \bibitem{Bat90} C.~J.~Batty,
	 Nucl. Phys. {\bf A 508}, 89c (1990).
	 
 \bibitem{Hir00} S.~Hirenzaki, Y.~Okumura,
	 H.~Toki, E.~Oset, and A.~Ramos,
	 Phys. Rev. {\bf C 61}, 055205 (2000).
	 
 \bibitem{FriPC} E.~Friedman, private communication.

 \bibitem{Sat07} M.~Sato {\em et al.}, to be submitted to Phys. Lett. B.
	 
 \bibitem{KETEK} {\tt http://www.ketek.net}

 \bibitem{NIST} R.~D.~Deslattes {\em et al.},
	 X-ray Transition Energies (version 1.2),
	 [Online] Available: http://physics.nist.gov/XrayTrans
	 [2007, May 16],
	 National Institute of Standards and Technology,
	 Gaithersburg, MD.

 \bibitem{XDB} A.~C.~Thompson {\em et al.}, X-ray data booklet,
	 LBNL/PUB-490 Rev. 2,
	 Lawrence Berkeley National Laboratory (2001).

 \bibitem{LECS}	R.~Marc Kippen, New Astronomy Reviews {\bf 48} (2004)
	 221-225 (2004).

 \bibitem{ResponseFunc01} J.~L.~Campbell, Leonard McDonald, Theodore
 Hopman and Tibor Papp, X-ray Spectrom. {\bf 30}, 230-241 (2001).
	 
 \bibitem{ResponseFunc03} M.~Van Gysel, P.~Lemberge and P.~Van Espen,
 X-ray Spectrom. {\bf 32}, 434-441 (2003).
	 
 \bibitem{KoiPC} T.~Koike, in preparation.

 \bibitem{PDG} W.-M.~Yao {\em et al.}, J. Phys. G {\bf 33}, 1 (2006).

 \bibitem{San05} J.~P.~Santos, F.~Parente, S.~Boucard,
	 P.~Indelicato, J.~P.~Desclaux,
	 Phys. Rev. {\bf A 71}, 032501 (2005).

 \bibitem{Mar06} J.~Mare$\check{\mbox{s}}$, E.~Friedman and A.~Gal,
	 Nucl. Phys. {\bf A 770}, 84 (2006).
	 
 \bibitem{Aka05} Y.~Akaishi, proceedings
	 for {\it International Conference on Exotic Atoms (EXA05)},
	 Austrian Academy of Sciences Press, Vienna, 2005, p. 45.
	 
 \bibitem{E17} R.~S.~Hayano {\em et al.},
	 Proposal of J-PARC 50-GeV PS
	 ``Precision spectroscopy of Kaonic Helium 3
	 $3d \to 2p$ X-rays'' (J-PARC E17) (2006).
	  
\end{thebibliography}
\end{document}